\documentclass[prl,twocolumn,superscriptaddress,showpacs,aps]{revtex4}

\usepackage{color}
\usepackage{graphicx}
\usepackage{bm}
\usepackage{amsmath}

\begin{document}

\newcommand{\ce}{CeCoIn$_5$}
\newcommand{\Tc}{T_c}
\newcommand{\Hup}{H_{c2}}
\newcommand{\Horb}{H_{\text{orb}}}
\newcommand{\fq}{\phi_{\circ}}
\newcommand{\qo}{q_{\circ}}
\newcommand{\xiorb}{\xi_{\text{orb}}}
\newcommand{\xiup}{\xi_{\text{c2}}}

\title{Field Dependent Coherence Length in the Superclean, High-$\kappa$
       Superconductor \ce}

\author{L. DeBeer-Schmitt}
\affiliation{Department of Physics, University of Notre Dame,
             Notre Dame, IN 46556 USA}

\author{C. D. Dewhurst}
\affiliation{Institut Laue-Langevin, 6 Rue Jules Horowitz,
             F-38042 Grenoble, France}

\author{B. W. Hoogenboom}
\affiliation{M.E. M\"uller Institute for Structural Biology, Biozentrum, and
             Institute of Physics, University of Basel,
             CH-4056 Basel, Switzerland}

\author{C. Petrovic}
\affiliation{Condensed Matter Physics and Materials Science Department,
             Brookhaven National Laboratory, Upton, NY 11973 USA}

\author{M. R. Eskildsen}
\email{eskildsen@nd.edu}
\affiliation{Department of Physics, University of Notre Dame,
             Notre Dame, IN 46556 USA}

\date{\today}
\begin{abstract}
Using small-angle neutron scattering, we have studied the flux-line lattice
(FLL) in superconducting \ce. The FLL is found to undergo a first-order
symmetry and reorientation transition at $\sim 0.55$ T at 50 mK. The FLL form
factor in this material is found to be independent of the applied magnetic
field, in striking contrast to the exponential decrease usually observed in
superconductors. This result is consistent with a strongly field-dependent
coherence length in \ce, in agreement with recent theoretical predictions for
superclean, high-$\kappa$ superconductors.
\end{abstract}


\maketitle

Since the discovery of heavy-fermion superconductivity in \ce, a plethora of
interesting phenomena have been observed in this material. Among these are one
of the highest critical temperature ($\Tc = 2.3$ K) in any heavy-fermion
superconductor \cite{Petrovic01}, $d$-wave pairing symmetry
\cite{Izawa01,Aoki04}, field- and pressure-induced quantum-critical points and
non-Fermi liquid behavior
\cite{Kim01,Sidorov02,Paglione03,Bianchi03a,Ronning05}, strong
indications of the first realization of a non-uniform
Fulde-Ferrell-Larkin-Ovchinnikov state
\cite{Murphy02,Radovan03,Bianchi03b,Won04,Martin05,Kakuyanagi05}, and
suggestions of multi-band or multi-order parameter superconductivity
\cite{Rourke05,Tanatar05}.

In addition, {\ce} is also found to represent an extreme case of a clean,
high-$\kappa$ superconductor. The elastic electronic mean free path in this
material ranges from $l$ = 840 {\AA} at $\Tc$ \cite{Movshovich01}, increasing
exponentially as the temperature is decreased and reaching  values of 4 to 5
$\mu$m at 400 mK \cite{Ormeno02,Park05}. Literature values for the penetration
depth varies from 190 nm \cite{Ormeno02} to 280 nm \cite{Ozcan03}. Estimates of
the orbital critical field based on $d\Hup/dT|_{\Tc}$ range from
$\Horb^{\|c}(0)= 13.2$ T \cite{Movshovich01} to 15.0 T \cite{Tayama02},
yielding an in-plane coherence length of $\xiorb = 47 - 50$ {\AA}, and hence
$\kappa$ = 40 - 60 and $l/\xi \sim 10^3$ or larger at temperatures below a few
hundred milliKelvin.

In many superconductors, detailed information about the nature of the
superconducting state has been obtained from small-angle neutron scattering
(SANS) studies of the flux-line lattice (FLL). Examples of this are FLL
symmetry transitions driven by non-local electrodynamics \cite{Eskildsen97} or
a superconducting gap anisotropy \cite{Huxley00,Brown04}, and subtle changes in
the fundamental length scales obtained from the FLL neutron reflectivity
\cite{Yaron97, Kealey00}. In this Letter we report on SANS studies of the FLL
in \ce. In striking contrast to the exponential dependence usually observed in
superconductors, the FLL form factor in this material is found to be
independent of the applied magnetic field. This result indicates a coherence
length which decreases with increasing field, in qualitative agreement with
recent theoretical predictions for clean, high-$\kappa$ superconductors
\cite{Kogan05}.

The SANS experiments were carried out at the D22 and D11 instruments at the
Institut Laue-Langevin. The {\ce} single crystals used in the experiment were
grown from excess indium flux, and had a $\Tc = 2.3$ K and a $\Hup(0) = 5.0$ T
for fields applied parallel to the $c$ axis \cite{Petrovic01}. The sample was
composed of three individually aligned single crystals with thicknesses 0.13 -
0.22 mm mounted side by side. The total mass of the sample was 86 mg. The use
of rather thin crystals was necessary, due to the strong neutron absorption by
indium. Incident neutrons with a wavelength of $\lambda_n$ = 4.5 {\AA} (D11)
and 7 {\AA} (D22) and a wavelength spread of
$\Delta\lambda_n/\lambda_n$ = 10$\%$ were used, and the FLL diffraction pattern
was collected by a position sensitive detector. For all measurements, the
sample was field-cooled to a base temperature of 40 - 50 mK in a dilution
refrigerator insert, placed in a superconducting cryomagnet. Horizontal
magnetic fields in the range 0.4 - 2.0 T were applied parallel to the
crystalline $c$ axis and the incoming neutrons. Background subtraction was
performed using measurements following zero-field cooling.

\begin{figure*}
  \includegraphics{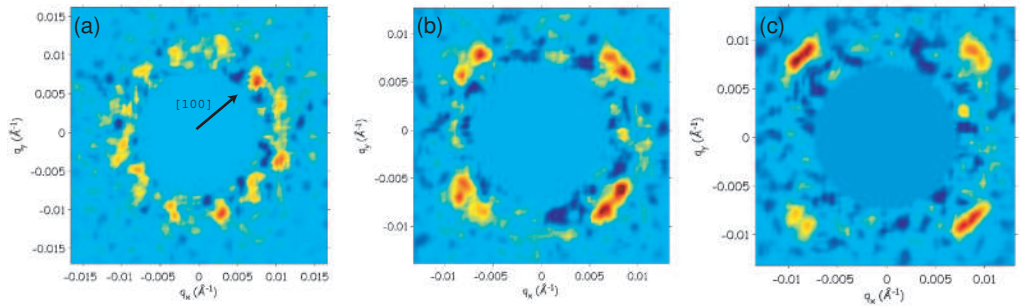}
  \caption{(Color) FLL diffraction patterns for {\ce} with applied fields
           0.5 T (a), 0.55 T (b) and 0.75 T (c), after subtraction of
           background measurement. The data is smoothed and the center of the
           image is masked off. The crystalline $a$ axis is vertical.
           \label{diffpat}}
\end{figure*}

Fig. \ref{diffpat} shows FLL diffraction patterns obtained at three
different applied fields.
Each image is a sum of the scattering from the FLL, as the sample is rotated
and tilted in order to satisfy the Bragg condition for the different
reflections. At fields below 0.55 T, 12 (2 $\times$ 6) reflections are observed
as shown in Fig. \ref{diffpat}(a), corresponding to two nearly hexagonal
domains with Bragg peaks aligned along $\langle 100 \rangle$-directions. As the
field is increased above $\sim 0.55$ T the FLL undergoes a first order
transition to a rhombic symmetry as shown in Figs. \ref{diffpat}(b) and (c).
Again two rhombic FLL domain orientations are observed, indicated by the 8
$(2 \times 4)$ Bragg peaks. As evident from the decreasing peak splitting in
the rhombic phase, the FLL gradually evolves towards a square symmetry as the
field is increased. These results are in agreement with our earlier studies
\cite{Eskildsen03}.

\begin{figure}[b]
  \includegraphics{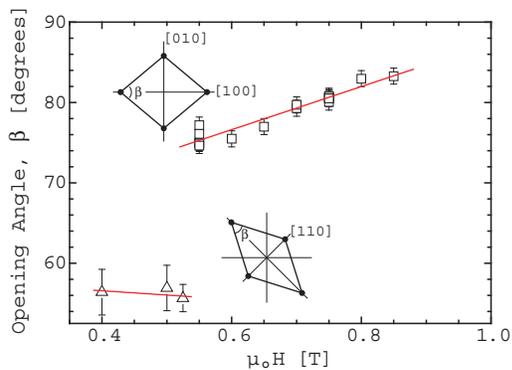}
  \caption{Field dependence of the FLL opening angle, $\beta$. The insets show
           the real space orientation of the FLL unit cell in the low and high
           field configurations.
           \label{beta}}
\end{figure}

The evolution of the symmetry transition, quantified by the FLL opening angle
$\beta$, is summarized in Fig. \ref{beta}.
Though it was not possible to reliably fit and extract a FLL split angle above
0.85 T, high-resolution measurements up to 1.0 T still showed a weak rhombic
distortion. A linear extrapolation of the opening angle to
$\beta = 90^{\circ}$, yields a transition field $H_2 = 1.1$ T above which a
perfectly square FLL is realized. Besides the opening angle, we also used
scattering vectors belonging to the same domain to obtain the internal field
by the relation $\bm{q}_1 \times \, \bm{q}_2 = (2\pi)^2 B/\fq$, where
$\fq = 20.7 \times 10^4$ T{\AA}$^2$ is the flux quantum. From this we obtain
$dB/d(\mu_{\circ}H) = 0.992 \pm 0.012$, which leads us to set
$B = \mu_{\circ}H$ for the remainder of this work.

A square FLL can either be stabilized by a gap anisotropy as observed e.g. in
YBCO \cite{Brown04}, or by non-local electrodynamics coupled with a Fermi
surface anisotropy as seen in the borocarbide superconductors
\cite{Eskildsen97,Kogan97}. In the case of {\ce} the orientation of the gap
nodes have been subject to controversy \cite{Izawa01,Aoki04,Tanatar05},
although recent theoretical work aimed at resolving this issue concluded that
the pairing symmetry in this material is $d_{x^2-y^2}$ \cite{Vorontsov06}. As
we have previously reported, such an orientation of the gap nodes is consistent
with the high-field square FLL being stabilized by $d$-wave pairing
\cite{Eskildsen03}. On the other hand, an extrapolation to $H = 0$ of the the
high-field opening angle in Fig. \ref{beta} yields $\beta \approx 60^{\circ}$,
as expected if the FLL symmetry is determined by non-local effects
\cite{Kogan97}.

\begin{figure}
  \includegraphics{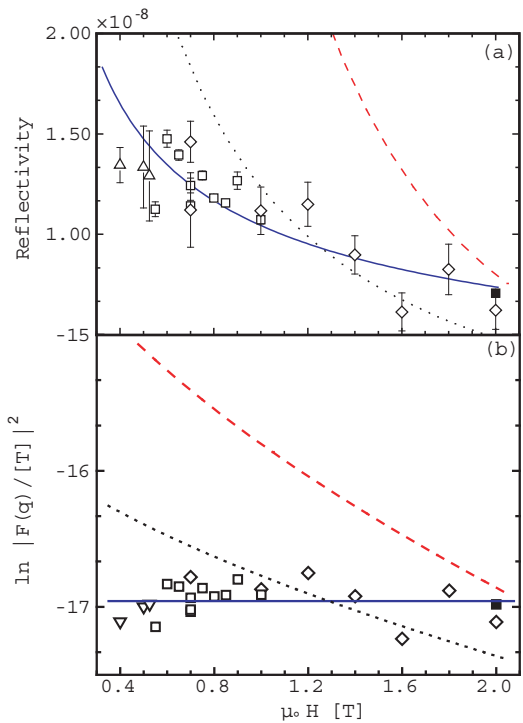}
  \caption{Field dependence of the reflectivity (a) and form factor (b) for the
          (1,0) FLL Bragg reflections. Triangles correspond to a hexagonal FLL
          and the squares and diamonds to a rhombic FLL. The solid square is
          from previous work \cite{Eskildsen03}. The dashed and dotted
          lines are calculated using Eqs. (\ref{refleq}) and (\ref{Clemeq})
          with the following values for the coherence length and the
          penetration depth:
          $\xi = 81$ {\AA}, $\lambda = 2350$ {\AA} (dashed line);
          $\xi = 50$ {\AA}, $\lambda = 3750$ {\AA} (dotted line). The solid
          line corresponds to a constant form factor,
          $F = 2.08  \times 10^{-4}$ T.
          \label{refl}}
\end{figure}

We now turn to measurements of the FLL form factor, which are the main results
of this Letter. Fig \ref{refl}(a) shows the FLL reflectivity obtained from
the integrated intensity of the Bragg peaks, as the sample is rotated through
the diffraction condition.
The reflectivity is given by
\begin{equation}
  R = \frac{2 \pi \gamma^{2} \lambda_n^2 t}{16 \fq^2 q} |F(q)|^2,
  \label{refleq}
\end{equation}
where $\gamma = 1.91$ is the neutron gyromagnetic ratio, $\lambda_n$ is the
neutron wavelength, $t$ is the sample thickness, and $q$ is the scattering
vector \cite{Christen77}. The form factor, $F(q)$, is the Fourier transform of
the magnetic field profile around a vortex, and depends on both the penetration
depth and the coherence length.
Fig. \ref{refl}(b) shows the FLL form factor determined from the reflectivity.
Here we have used the measured opening angle, $\beta(H)$, in Fig. \ref{beta} to
determine the magnitude of the scattering vector, and hence compensate for
effects due to the FLL symmetry transition. In the high-field rhombic FLL phase
$q_{\text{sq}} = \qo / \sqrt{\sin \beta}$ where $\qo = 2\pi \sqrt{B/\fq}$. In
the low-field distorted hexagonal phase, the 4 Bragg peaks aligned along the
$\langle 100 \rangle$-direction have
$q_{\text{hex1}} = \qo \sqrt{2(1 - \cos \beta)/\sin \beta}$, while for the
remaining 8 peaks $q_{\text{hex2}} = \qo /\sqrt{\sin \beta}$.

The field-independent form factor observed for {\ce} is in striking contrast to
the exponential decrease observed in other superconductors. However, it is
important to note that if the form factor was truly independent of $q$, this
would imply an unphysical situation with a diverging magnetic field at the
vortex center and a coherence length equal to zero. To reconcile this apparent
paradox, it is necessary to assume a field dependence of either the
superconducting penetration depth, the coherence length or both.

Several models exists for the magnetic field profile around the vortices. In
the following we have based the analysis on the form factor obtained by Clem
\cite{Clem75}:
\begin{equation}
  F(q) = B \frac{g \; K_{1}(g)}{1 + \lambda^2 q^2}, \hspace {1cm}
  g = \sqrt{2} \xi \; (q^2 + \lambda^{-2})^{1/2},
  \label{Clemeq}
\end{equation}
valid for $\kappa \gg 1$. It is worth to point out that while this form factor
was obtained for an isotropic superconductor, numerical simulations found no
significant difference between superconductors with $s$- or $d$-wave pairing
\cite{Ichioka99}.

The dashed lines in Figs. \ref{refl}(a) and (b) show the reflectivity and form
factor calculated from Eqs. (\ref{refleq}) and (\ref{Clemeq}) using an average
literature value for the penetration depth $\lambda = 2350$ {\AA} and
$\xiup = \sqrt{\frac{\fq}{2 \pi \Hup}} = 81$ {\AA}. As already discussed, these
calculated values are not consistent with the experimental data. However, a
more appropriate value for the coherence length is obtained from the orbital
critical field, $\xiorb = \sqrt{\frac{\fq}{2 \pi \Horb}} \approx 50$ {\AA}
\cite{Movshovich01,Tayama02}. This does correspond to a smaller slope of
$|F(q)|^2$, but also increases the discrepancy between the measured and
calculated values of $|F(q)|^2$ since, for a given  $\xi$ and $q$ (or $H$), the
magnitude of the form factor is determined by $\lambda^2$. Using the
penetration depth as a fitting parameter, one obtains the dotted lines in
Fig. \ref{refl} for $\lambda = 3750$ {\AA}. Although this provides a better
agreement with the data, such a large value of $\lambda$ is not consistent with
reports in the literature \cite{Ormeno02,Ozcan03}. In contrast, a perfect fit
to the data is obtained with a constant value of the form factor,
$F = 2.08 \times 10^{-4}$ T, and correspondingly a reflectivity $\propto 1/q$,
as shown by the solid lines in Fig. \ref{refl}.

It is important to note that no significant disordering of the FLL is observed.
Except for systematic differences between the two beamlines, the FLL rocking
curve widths remain essentially constant throughout the measured field range.
On D22 we find rocking curve widths going from $0.14^{\circ} \pm 0.02^{\circ}$
FWHM at 0.55 T to $0.16^{\circ} \pm 0.01^{\circ}$ FWHM at 1 T, comparable to 
the experimental resolution ($\sim 0.08^{\circ}$ FWHM). Below the
reorientation transition a slightly higher value of
$0.20^{\circ} \pm 0.04^{\circ}$ FWHM is observed. On D11 FLL rocking curve
widths decrease from $0.26^{\circ} \pm 0.03^{\circ}$ FWHM at 0.7 T to
$0.19^{\circ} \pm 0.01^{\circ}$ FWHM at 2 T.  Such narrow rocking curve widths
indicate a very well ordered FLL with a longitudinal correlation length in the
micron range, consistent with weak pinning due to the high cleanliness of \ce.
Furthermore, FLL disorder above a certain threshold has been shown to lead to
decrease in the scattered intensity, exceeding the usual exponential field
dependence of the form factor \cite{Klein01}. We can therefore exclude FLL
disordering as the cause for the field-independence of the form factor in \ce.

A constant form factor could in principle be due to a decrease of the
penetration depth with increasing field, caused either by an increasing
superfluid density or a non-uniform spin magnetization contributing to the
magnetic flux carried by each vortex. We do not consider the first possibility
realistic. Furthermore, while {\ce} is indeed paramagnetic \cite{Tayama02},
spin polarization effects will lead to an enhancement of the form factor in
contrast to the strong suppression observed in this material
\cite{Tachiki79,Machida06}. We therefore conclude that while paramagnetic
effects may contribute, they are not the dominating mechanism behind the
field-independent form factor. In the following we therefore restrict our
analysis to consider only a field-dependent coherence length.

\begin{figure}
  \includegraphics{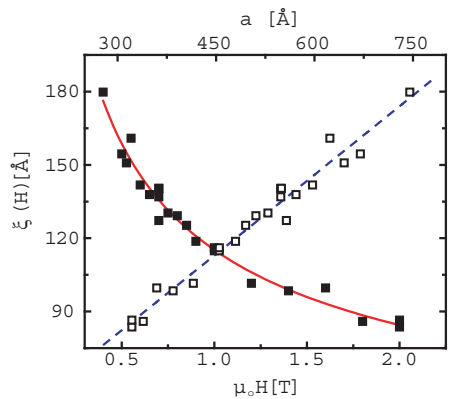}
  \caption{Field dependence of the coherence length, $\xi(H)$ plotted versus
           applied field (solid symbols), and vortex separation (open symbols).
           The solid and dashed lines are fits to the data described in the
           text.
           \label{xi}}
\end{figure}

In Fig. \ref{xi}, we show the coherence length obtained by varying
$\xi$ to achieve the measured form factor at each field.
The coherence length is found to follow a $1/\sqrt{H}$ behavior. While
different models for $F(q)$ will provide slightly different values for the
coherence length, the qualitative behavior will remain unchanged.
Fig. \ref{xi} also shows the extracted coherence length as a function of
intervortex spacing, $a$. Within the experimental error shown by the scatter
in the data, the coherence length is found to increase linearly with the
intervortex spacing while at all times satisfying $\xi < a$. Extrapolating to a
vortex separation of 125 {\AA} (corresponding to an orbital critical field
$\Horb = 13.2$ T) yields $\xi= 43.0 \pm 4.6$ {\AA}, in reasonable agreement
with $\xiorb = 50$ {\AA}.

We believe that our data provide the first example of a strongly
field-dependent coherence length in a superclean, high-$\kappa$ superconductor
as recently theoretically predicted by Kogan and Zhelezina \cite{Kogan05}. In
their model, the coherence length is predicted to be proportional to
$1/\sqrt{H}$ and correspondingly depends linearly on the vortex separation.
Experimentally we find $d\xi/da = (0.55 \pm 0.02)/\sqrt{2\pi}$, which is
roughly twice the theoretically predicted slope \cite{Kogan05}. However, the
theoretical prediction was based on assumptions of weak-coupling $s$-wave
superconductivity and a simple Fermi surface topology (sphere or cylinder),
which are not valid for \ce. Based on this we argue that the agreement between
theory and experiment is remarkable, supporting the universal nature of the
effect. At low fields the model predicts the coherence length to saturate.
Based on our experimental results this cut-off in {\ce} occur below the lowest
measured field of 0.4 T. We believe that the field-dependent coherence length
is so prominently observed in {\ce} due to the combination of a large $\kappa$
and the very high cleanliness of this material.

Additional evidence for the unusual magnetic-field response of {\ce} has been
observed in the quasiparticle mean free path, as extracted from measurements of
the Hall angle \cite{Kasahara05}. Specifically, the mean free path is found to
decrease as the applied field increases, being roughly equal to the vortex
separation for the range of fields covered by our SANS measurements. This leads
us to speculate that a connection exists between $\xi$ and $l$ beyond the
simple Pippard model.

Finally, we want to note that while previously reports of a field dependent
vortex core size have been made by Sonier \textit{et al.}
\cite{Sonier04,Callaghan05}, these were attributed to a vortex-squeezing
effect. However, since the variation of the vortex core size in these cases was
significantly smaller than for \ce, we do not believe that such effects can
explain the data reported here.

In summary, we have presented SANS measurements of the FLL in {\ce}. These
measurements indicate a  strong reduction of the superconducting coherence
length with increasing field, in qualitative agreement with recent theoretical
predictions for superclean, high-$\kappa$ superconductors.

\begin{acknowledgments}
We are grateful to N. Jenkins for assistance with the SANS measurements and to
K. Machida for valuable discussions.
MRE acknowledges support by the Alfred P. Sloan Foundation, and BWH support by
the NCCR Nanoscale Science.
Part of this work was carried out at Brookhaven National Laboratory which is
operated for the US Department of Energy by Brookhaven Science Associates
(DE-Ac02-98CH10886).
\end{acknowledgments}


\newpage

\printfigures

\end{document}